\documentclass[twocolumn,showpacs,prl]{revtex4}

\usepackage{graphicx}

\begin{document}

\title{Laser-driven nanoplasmas in doped helium droplets:\\
  Local ignition and anisotropic growth}

\author{Alexey Mikaberidze}
\author{Ulf Saalmann}
\author{Jan M. Rost}
\affiliation{Max Planck Institute for the Physics of Complex
  Systems, N\"{o}thnitzer Stra{\ss}e 38, 01187 Dresden, Germany}

\begin{abstract}
\noindent Doping a helium nanodroplet with a tiny xenon cluster of a
few atoms only, sparks complete ionization of the droplet at laser
intensities below the ionization threshold of helium atoms.  As a
result, the intrinsically inert and transparent droplet turns into a
fast and strong absorber of infrared light. Microscopic calculations
reveal a two-step mechanism to be responsible for the dramatic change:
Avalanche-like ionization of the helium atoms on a femtosecond time
scale, driven by field ionization due to the quickly charged xenon
core is followed by resonant absorption enabled by an unusual
cigar-shaped nanoplasma within the droplet.
\end{abstract}


\maketitle
\noindent
Helium nanodroplets are fascinating objects due to their multi-facet
properties, ranging from superfluidity itself \cite{grto+98} to
unusual electron dynamics upon photon impact \cite{peli+03}.  Most
widely, however, helium droplets are used as a ``catalyst'' to
facilitate an intended physical process without altering it.  
A prime example is the cooling of molecules inside a
droplet for high-resolution spectroscopy \cite{tovi04,stle06}, the
creation of unusual molecules on the droplet surface \cite{hica+96},
but also the assembly and transport of clusters inside a helium
droplet \cite{mosl+07}.  The catalytic property originates in the high
ionization potential of helium in connection with the almost
frictionless, superfluid environment.

Here, we will show that also the opposite is possible: By doping
the droplet with a handful of heavier rare-gas atoms the inert
and transparent helium droplet can be turned into a highly 
reactive object which absorbs infrared light very effectively.  
In this case, the heavier atoms serve as a ``catalyst'' to
activate the helium droplet.
This finding is quite surprising since the pristine droplet
cannot be ionized at all with laser light of 780\,nm wavelength
at an intensity of $I \sim 10^{14}$\,W/cm$^2$ which we will
apply.  
Yet, with a few xenon
atoms inside, {\em all} electrons from the helium atoms are removed so
that the entire droplet containing as many as $10^{5}$ helium atoms
turns into a nanoplasma.  The extremely efficient ionization -- even for
large droplets -- requires two elements, an initial seed and a resonant
energy absorption process from the laser pulse. 
The latter occurs on an \emph{electronic\/} time scale, in
contrast to the well known resonant absorption during the
Coulomb explosion of a homogenous cluster which occurs on the
much slower time scale of \emph{nuclear\/} motion
\cite{dido+96,saro03sa06}. 

In order to take the role of a seed the ionization potential of
the embedded species must be lower than that of helium
($E_\mathrm{ip}{=}24.6$\,eV).  We have used xenon
($E_\mathrm{ip}{=}12.1$\,eV) for the present investigation. 
As known from quantum Monte Carlo calculations \cite{bole+08},
rare-gas dopants settle in the droplet center.
When the laser pulse is ramped up, xenon is ionized first
producing a strong static electric field, which is capable to
significantly ionize helium in combination with the laser
field.  
In this first phase, electrons
from the droplet migrate to the center of the cluster, but also leave
the cluster, i.\,e., are outer ionized, increasing the total cluster
charge and thereby enhancing helium ionization.  However, the degree
of ionization per helium atom, which can be achieved with field
ionization, decreases rapidly with the helium droplet size, since the
static field of the ionic charge falls off quadratically with the
distance from the center.

Hence, to turn very large droplets into completely stripped atoms,
another, even more powerful mechanism for ionization has to kick in
and this is a new kind of plasma resonance which leads to an
ionization avalanche.  A little thought reveals that resonant plasma
absorption should not be possible because the dipole plasmon frequency
$\Omega$ of a spherical helium nanoplasma is too high to come into
resonance with a short laser pulse of frequency $\omega$.  However,
since the ionization starts from a small seed at the droplet center
and is driven by a linearly polarized laser, a non-spherical,
cigar-shaped nanoplasma forms.  It has a plasmon frequency along the
longer axis \emph{lower\/} than that of a sphere with an equal charge
density.  Therefore, resonance with the laser frequency is possible
and happens in fact very fast, within a few femtoseconds.  
As a result, helium droplets containing more than $10^5$ atoms
are completely ionized with the help of 13 or fewer embedded
xenon atoms, while in the absence of the xenon seed, the droplet is not ionized at
all.  The ionization avalanche can be invoked by a wide range of laser
frequencies, intensities and pulse durations.  It is also robust with
respect to the helium density and the species of seed atoms as long as
they have lower ionization potential than helium.

\begin{figure}[t]
\centerline{\includegraphics[width=0.7\columnwidth]{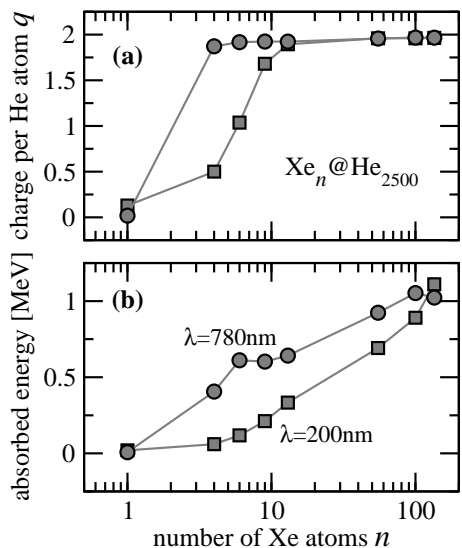}} 
\caption{Ionization and energy absorption of a He$_{2500}$
  droplet doped with a Xe$_n$ cluster.
  (a) Charge per helium atom and (b) absorbed energy
  as a function of the number of xenon atoms $n$.
  The charge $q$ refers to the average number of electrons
  released from a droplet atom. 
  Both quantities are shown for two different laser wavelengths:
  $\lambda = 780$\,nm (circles) and $\lambda = 200$\,nm
  (squares). 
  The other laser parameters were identical:
  peak intensity $I=7\times10^{14}$\,W/cm$^2$,
  Gaussian pulse $\exp(-\ln2(t/T)^2)$ with a duration
  $T=20$\,fs, linear polarization.}  
\label{fig:ndep}
\end{figure}%
In order to demonstrate the two elements necessary for the avalanche
ionization and in particular the new anisotropic nanoplasma resonance,
we will present calculations for $\lambda=780$\,nm and for
$\lambda=200$\,nm laser light, because in the latter case the laser
frequency is too high to be matched by an eigenfrequency of the
anisotropic nanoplasma and only the first mechanism is active,
namely field ionization of the helium droplet.
We use a classical molecular dynamics approach 
described elsewhere \cite{misa+08},
which we have developed previously 
\cite{saro03sa06,sasi+06} and which has been applied in a similar
way also by other authors \cite{isbl00,lajo01,fera+07}.  For droplets
with more than $2500$ atoms we have used an implementation
\cite{daka08} of the fast-multipole method \cite{grro87} to calculate
the Coulomb interaction of electrons and ions.

\begin{figure}[b]
\centerline{\includegraphics[width=0.7\columnwidth]{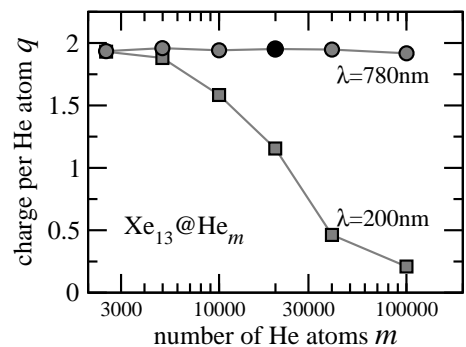}}
  \caption{Charge per helium atom as in Fig.\,\ref{fig:ndep}a for
  helium droplets doped with a Xe$_{13}$ core as a function of the
  droplet size $m$.  The laser parameters are the same as in
  Fig.\,\ref{fig:ndep}.
  The black circle marks the droplet shown in
  Fig.\,\ref{fig:extension}.}
  \label{fig:mdep}
\end{figure}

Firstly, we will show that the avalanche ionization effect occurs over a wide
range of sizes of embedded xenon cluster $n$ -- starting from a very few
atoms --  and also over a wide range of helium droplet sizes $m$, spanning
more than two orders of magnitude (Figs.\,\ref{fig:ndep} and
\ref{fig:mdep}).  Secondly, we will analyze the ionization dynamics of a
particular cluster to understand the peculiarities of the resonance 
mechanism here at work.

Figure~\ref{fig:ndep} summarizes the dependence of the avalanche
ionization on the xenon core size $n$ for a fixed droplet containing
$m=2500$ helium atoms.  As can be seen in Fig.\,\ref{fig:ndep}, already
very few embedded xenon atoms lead to complete inner ionization,
whereby inner ionization refers to the (average) number $q$ of
electrons released from a droplet atom,
with the maximum value $q=2$ for helium.
Of course, these electrons may still be 
trapped by the global droplet potential forming a plasma as 
discussed below.  Outer ionization, i.\,e., the removal of electrons
from the entire cluster, is determined by the plasma dynamics and may
depend on post-pulse effects \cite{fera+07}.

The average charge of two indicates that the entire  droplet has
 turned into a nanoplasma, i.\,e.\ there are neither neutral
atoms nor singly charged helium ions left.  How dramatic this effect is,
becomes also evident in the absorbed energy, cf.\
Fig.\,\ref{fig:ndep}b.  Whereas the absorption for one xenon atom is
negligible, it reaches for cores with $n=4\ldots13$ xenon atoms a few
hundred keV due to the ignition of the droplet.  The increase occurs more
abruptly for the longer wavelength ($\lambda=780$\,nm, circles in
Fig.\,\ref{fig:ndep}) than the shorter one ($\lambda=200$\,nm, squares
in Fig.\,\ref{fig:ndep}) as can be seen for the charge $q$ in panel
(a) as well as for the absorbed energy in panel (b).  The energy
increases further for cores with $n=55\ldots135$ xenon atoms up to values of
about 1\,MeV, whereby now the core itself contributes considerably to
the absorption.

The difference in absorption between the two laser wavelengths 
becomes much clearer from
the dependence on the helium droplet size $m$ in
Fig.\,\ref{fig:mdep},  where the size of the core Xe$_{13}$ was kept
constant.  Increasing
the droplet size $m$ by almost two orders of magnitude does not change
the behavior for the larger wavelength (circles in
Fig.\,\ref{fig:mdep}): Up to a size of $m\sim10^5$ we find complete
inner ionization.  We recall  that not a single of these helium atoms
would be ionized without the xenon core.  In contrast, at shorter
wavelength the charge per atom decreases for larger droplets (squares in
Fig.\,\ref{fig:mdep}) since the static field generated by the xenon 
core ions can  only ionize a droplet of a certain size as 
the field strength drops quadratically with
the distance.  Obviously, the ratio of xenon to helium
atoms is crucial.  For smaller droplets $m\ll10^4$
the core ions alone drive the complete helium ionization, for larger
droplets $m\gg10^4$ they do not.  
The reason for the different behavior at the two frequencies is
resonant absorption.  We will discuss it in detail since it exhibits
novel features which do not occur in homo-nuclear clusters.

\begin{figure}[t]
\centerline{\includegraphics[width=0.7\columnwidth]{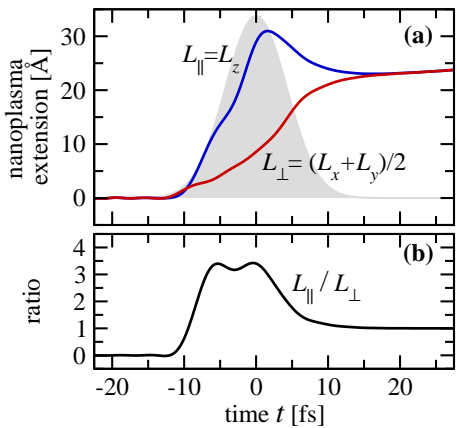}}
\caption{(color online).
  Anisotropic nanoplasma formation.
  (a) Extensions $L_\parallel$ and $L_\perp$ of the nanoplasma 
  parallel (blue/upper line) and perpendicular (red/lower) to
  the laser polarization as a function of time $t$
  for the droplet marked in Fig.\,\ref{fig:mdep} ($m\,{=}\,20000$).
  Lines are averaged over a laser period in order to smooth 
  sub-cycle structures.
  The gray-shaded area sketches the laser pulse envelope.
  (b) Ratio $L_\parallel/L_\perp$ as obtained for the smoothed
  curves from panel a.}
\label{fig:extension}
\end{figure}%
Helium's nuclear charge of two does not allow for charge
densities $\varrho$ larger than two times the particle density of the
droplet: $\varrho_\mathrm{max}=0.04/$\AA$^{3}$.  
The corresponding eigenfrequency for a dipole oscillation of an 
electron cloud with respect to the ionic background, that has a
uniform charge density $\varrho$, reads for spherical geometry
$\Omega=\Omega_\mathrm{pl}/\sqrt{3}$ with the plasma frequency
$\Omega_\mathrm{pl}=\sqrt{4\pi\varrho}$ \cite[Sect.\,2.1.2]{krvo98}
resulting in $\Omega_\mathrm{max}=0.16$\,a.u.\ for a completely
inner-ionized helium droplet.  
With this eigenfrequency it is impossible to match
the laser frequency $\omega=0.23$\,a.u.\ for the shorter wavelength
$\lambda=200$\,nm.  On the other hand, $\Omega_\mathrm{max}$ is
considerably larger than $\omega=0.058$\,a.u., the laser frequency for
$\lambda=780$\,nm.  To reach the resonance, the eigenfrequency
of the droplet could be lowered 
by expansion of the sphere, which is the case for the usual nanoplasma
resonance \cite{dido+96,saro03sa06}.  This, however, can be ruled out
here, since we observe resonant absorption for a pulse length of
20\,fs which is too short for sufficient droplet expansion, even for
the light helium nuclei.  In fact, the nuclei are almost frozen during
the pulse.  Therefore, the eigenfrequency of the droplet must be
lowered by a different mechanism to come into resonance with the laser
frequency.

\begin{figure}[b]
\centerline{\includegraphics[width=0.9\columnwidth]{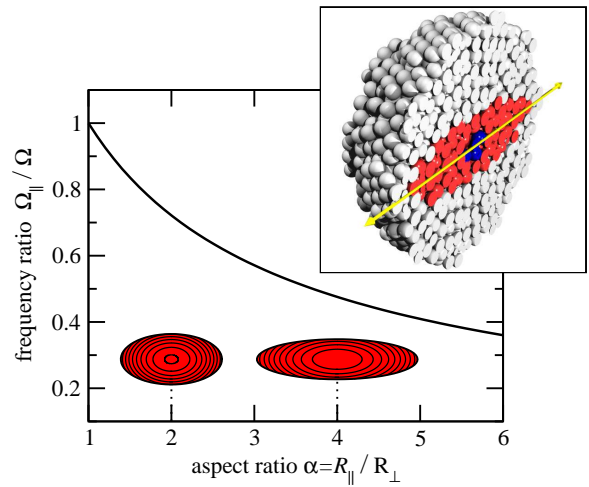}}
\caption{(color online).
  Elongated nanoplasma geometry. 
  Eigenfrequency $\Omega_\parallel$ for an ellipsoidal nanoplasma
  in the direction of the long axis $z$ in terms of the
  eigenfrequency for a spherical nanoplasma $\Omega$ as a
  function of its aspect ratio $\alpha=R_\parallel/R_\perp$
  according to Eq.\,(\ref{eq:efaz}). 
  Upper inset: Cross section of the nanoplasma (red/gray atoms)
  around the xenon core (blue/darkgray atoms) in the helium
  droplet (lightgray atoms).
  The arrow marks the laser polarization.
  Lower insets: Cross sections of equipotential surfaces in
  homogeneously charged ellipsoids with aspect ratios of
  $\alpha{=}2$ and $\alpha{=}4$, respectively.}
\label{fig:eigfreq:rat}
\end{figure}%
The key to understand this mechanism  is the \emph{geometry} of the 
nanoplasma formed.
 Our calculations reveal that ionization starts from the doped core in the
center of the droplet and subsequently spreads towards the droplet surface.
Most importantly, this spreading occurs anisotropically, namely
faster along the laser polarization than perpendicular to it, 
as shown in Fig.\,\ref{fig:extension}. 
There, we have plotted the 3-dimensional extension of the
nanoplasma using
\begin{equation}
  \label{eq:npex}
  L_x(t)=\sum_{i=1}^{n+m} q_i(t)\left|x_i(t)\right|
  \bigg/\sum_{i=1}^{n+m}q_i(t)
\end{equation}
and corresponding definitions for $L_y$ and $L_z$,
where $q_i(t)$ is the charge state of the $i$th atom located at
$\vec r_{i}=(x_i, y_i,z_i)$.  
The sum runs over all atoms.
Since our system is axially symmetric for a laser
linearly polarized in $z$-direction, we plot the extension 
$L_{\parallel}=L_{z}$ along the laser polarization $\hat z$ and 
perpendicular to it, $L_\perp=(L_x{+}L_y)/2$.
Figure~\ref{fig:extension}a shows clearly that  the
nanoplasma spreads much faster along the laser polarization
axis  with little extension perpendicularly to it
until the peak of the laser pulse at $t=0$  is reached. 
After the laser peak for times $t>0$ the plasma grows with
increased rate perpendicularly to the polarization.
At the same time, the extension along the polarization axis  
decreases since more atoms close to the ``equatorial plane''
with $z\approx0$ are ionized.
From about $15$\,fs on the whole droplet is ionized giving rise to a 
 spherical nanoplasma for which  $L_\perp=L_\parallel$.
At this time all helium ions are charged and the 
extensions characterize the droplet size.
The slight increase of both values for $t>15$\,fs reflects
the onset of droplet expansion.
In short, the ionization starts from the few xenon atoms in the
center of the droplet forming an axially symmetric plasma
channel.   
This plasma channel gradually elongates, cf.\ upper inset of
Fig.\,\ref{fig:eigfreq:rat},  and reaches the surface of the
droplet.   Only after that, the ionization spreads into the
whole droplet. 

To get an estimate on the plasma frequencies involved, we approximate the
elongated plasma channel by a cigar-shaped axially
symmetric ellipsoid with an extension $R_\parallel=R\alpha^{2/3}$ along
the laser polarization and a radius $R_\perp=R/\alpha^{1/3}$
perpendicular to it. 
Hereby, $R$ is the radius of a corresponding spherical
nanoplasma with the same volume and $\alpha =R_\parallel/R_\perp\ge1$ the
aspect ratio of the ellipsoid axes.
The potential inside a homogeneously charged ellipsoid 
follows from the general integral expression
\cite[\S\,99]{lali80} and reads 
\begin{equation} \label{eq:poel}
\phi(\rho,z;\alpha) = \pi \varrho 
\Big[\left[ 1{-}g(\alpha) \right] \rho^2  
+ 2 g(\alpha) z^2 - h(\alpha)R^2 \Big],
\end{equation}
with $g(\alpha)=\left[\alpha\ln\left(\alpha{+}f(\alpha)\right)/f(\alpha)
-1\right]/f^2(\alpha)$,
$h(\alpha)=\left[1{+}f^2(\alpha)g(\alpha)\right]/\alpha^{2/3}$,
and $f(\alpha)=\sqrt{\alpha^2{-}1}$.
In the limit of equal axes $R_\parallel=R_\perp$ the potential
Eq.~(\ref{eq:poel}) reduces to that of a sphere
$\phi(\rho,z;\alpha{\to}1) =  2\pi 
\varrho\left[\left(\rho^2{+}z^2\right)/3-R^2\right]$.
Note that the potential Eq.\,(\ref{eq:poel}) is harmonic in
$\rho$ and $z$ with ellipsoids as equipotential surfaces shown in
the lower insets of Fig.\,\ref{fig:eigfreq:rat}. 
However, the shape of these ellipses 
does not coincide with the plasma shape, and therefore the 
ratio of the eigenfrequencies differs from the aspect ratio of the 
nanoplasma.
The separation of $\rho$ and $z$ in Eq.\,(\ref{eq:poel}) makes
it straight-forward to calculate the eigenfrequency of a
collective dipole oscillation along the $z$ axis, i.\,e.\ along
the laser polarization.
We get for this frequency
$\Omega_\parallel{}^2(\alpha)=\frac{\partial^2}{\partial
  z^2}\phi(\rho,z;\alpha)\big|_{\rho,z=0}$  
from Eq.\,(\ref{eq:poel})  
in terms of the spherical eigenfrequency $\Omega$ defined above
\begin{equation}
 \label{eq:efaz}
  \Omega_\parallel(\alpha)/\Omega = \sqrt{3g(\alpha)}\,,
\end{equation}
which is shown in Fig.\,\ref{fig:eigfreq:rat}.  For all values
$\alpha>1$ the eigenfrequency is lower than $\Omega$, with a
reduction of almost 50\,\% for an aspect ratio of $\alpha=3$.

This explains the resonant energy absorption: As the plasma channel
elongates due to the inner ionization along the polarization
direction, the aspect ratio grows and the eigenfrequency
$\Omega_{\parallel}$ decreases.  At aspect ratios of
$\alpha\approx3\ldots4$, cf.\ Fig.\,\ref{fig:extension}b, the
ellipsoidal eigenfrequency $\Omega_{\parallel}$ is two times smaller
than the corresponding eigenfrequency $\Omega$ of a sphere.  Around
this aspect ratio, $\Omega_{\parallel}$ coincides with the laser
frequency giving rise to the plasma resonance.  As a consequence,
enough laser energy is absorbed by the plasma electrons, that the
ionization continues in the perpendicular direction and encompasses
the whole droplet.  This happens very fast, since the resonance
condition in form of the suitable aspect ratio of the nanoplasma is
created by an electronic process.  It does not rely on the slower
nuclear motion of an exploding droplet as it is the case in the
conventional resonant heating of a cluster.

To summarize, we have shown that helium nano\-droplets doped with
small xenon clusters turn into nanometer-sized plasmas with all helium
electrons removed.  This can be achieved with a laser pulse of
moderate intensity for which the pristine helium droplet is fully
transparent.  We observe an ionization avalanche on a fast timescale
of a few femtoseconds.  The phenomenon is robust over a wide range of
laser pulse parameters
and could be experimentally seen by detecting the depletion of
droplets from the beam or by measuring the abundance of helium ions
provided the droplets were doped by small xenon clusters. 

While we have discussed it from the perspective of the helium
nanodroplet, one can also view it in a more general context of
composite clusters, where interesting electron dynamics has been
observed for so called core shell systems (Ar-Xe clusters) under
ultraviolet laser pulses \cite{hobo+08}.  
In another experiment, a strong enhancement of X-rays from
laser-irradiated argon clusters doped by a few percent of water
molecules was observed \cite{jhma+05}.  
We have indications that this can be explained by a seed cluster
similarly as in the present context of the helium droplet
\cite{misa+08x}. 
In another scenario, deuterons, fast enough to induce nuclear
fusion \cite{dizw+99} can be generated in hetero-nuclear
clusters \cite{lajo01,hosy+05}, certainly also spectacular as
the complete stripping of helium in large nanodroplets sparked
by a few xenon atoms as discussed here.

Finally, the ionization avalanche effect in non-spherical plasmas
which we have discovered may be of interest far beyond cluster
physics.  Similar phenomena should also occur for laser-illuminated
solids doped with easy-ionizable impurities.  This could be a way to
create and study microscopic plasmas \emph{inside\/} solids
\cite{gama08}.


\begin{thebibliography}{10}

\bibitem{grto+98}
S. Grebenev, J.~P. Toennies, and A.~F. Vilesov, Science {\bf 279},  2083
  (1998).

\bibitem{peli+03}
D.~S. Peterka, A. Lindinger, L. Poisson, M. Ahmed, and D.~M. Neumark, 
Phys. Rev. Lett. {\bf 91},  043401  (2003).

\bibitem{tovi04}
J.~P. Toennies and A.~F. Vilesov, Angew. Chem. Int. Ed. {\bf 43},  2622
  (2004).

\bibitem{stle06}
F. Stienkemeier and K.~K. Lehmann, J. Phys. B {\bf 39},  R\,127  (2006).

\bibitem{hica+96}
J. Higgins {\it et~al.}, Science {\bf 273},  629  (1996).

\bibitem{mosl+07}
V. Mozhayskiy, M.~N. Slipchenko, V.~K. Adamchuk, and A.~F. Vilesov, J. Chem.
  Phys. {\bf 127},  094701  (2007).

\bibitem{dido+96}
T. Ditmire, T. Donnelly, A.~M. Rubenchik, R.~W. Falcone, and M.~D. Perry, 
Phys. Rev. A {\bf 53},  3379  (1996).

\bibitem{saro03sa06}
U. Saalmann and J.~M. Rost, Phys. Rev. Lett. {\bf 91},  223401  (2003);
U. Saalmann, J. Mod. Opt. {\bf 53},  173  (2006).

\bibitem{bole+08}
D. Bonhommeau, M. Lewerenz, and N. Halberstadt, J. Chem. Phys. {\bf 128},
  054302  (2008).

\bibitem{misa+08}
A. Mikaberidze, U. Saalmann, and J.~M. Rost, Phys. Rev. A {\bf 77},  041201(R)
  (2008).

\bibitem{sasi+06}
U. Saalmann, C. Siedschlag, and J.~M. Rost, J. Phys. B {\bf 39},  R\,39
  (2006).

\bibitem{isbl00}
K. Ishikawa and T. Blenski, Phys. Rev. A {\bf 62},  063204  (2000).

\bibitem{lajo01}
I. Last and J. Jortner, Phys. Rev. Lett. {\bf 87},  033401  (2001).

\bibitem{fera+07}
T. Fennel, L. Ramunno, and T. Brabec, Phys. Rev. Lett. {\bf 99},  233401
  (2007).

\bibitem{daka08}
H. Dachsel and I. Kabadshow,\\ http://www.fz-juelich.de/jsc/fmm, (2008).

\bibitem{grro87}
L. Greengard and V. Rokhlin, J. Comput. Phys. {\bf 73},  325  (1987).

\bibitem{krvo98}
U. Kreibig and M. Vollmer, {\em Optical Properties of Metal Clusters} 
(Springer Heidelberg, 1998).

\bibitem{lali80}
L.~D. Landau and E.~M. Lifschitz, {\em The Classical Theory of Fields}
  (Butterworth-Heinemann, Oxford, 1980).

\bibitem{hobo+08}
M. Hoener {\it et~al.}, J. Phys. B {\bf 41},  181001  (2008).

\bibitem{jhma+05}
J. Jha, D. Mathur, and M. Krishnamurthy, J. Phys. B {\bf 38},  L\,291  (2005).

\bibitem{misa+08x}
A. Mikaberidze, U. Saalmann, and J.~M. Rost, unpublished, 2008.

\bibitem{dizw+99}
T. Ditmire {\it et~al.}, Nature {\bf 398},  489  (1999).

\bibitem{hosy+05}
M. Hohenberger {\it et~al.}, Phys. Rev. Lett. {\bf 95},  195003  (2005).

\bibitem{gama08}
R.~R. Gattass and E. Mazur, Nat. Photon. {\bf 2},  219  (2008).

\end{thebibliography}
\end{document}